# Entanglement of Two Quantum Dots with the Flip-Flop Interaction coupled to Plasmonic Waveguide

Myong-Chol Ko,[1] Nam-Chol Kim,[1,*] Nam-Chol Ho,[2] Jian-Bo Li,[3]

[1]Faculty of Physics, **Kim Il Sung** University, Pyongyang, D P R of Korea

[2]Faculty of Physics, University of Science, Pyongyang, D P R of Korea

[3]Institute of Mathematics and Physics, Central South University of Forestry and Technology, Changsha 410004, China

[*]ryongnam10@yahoo.com,

**Abstract:** We investigate theoretically the entanglement of two quantum dots (QDs) coupled to metallic nanowaveguide in the presence of the flip-flop interaction with the analytical solutions of eigenvalue equations of the coupled system. High entanglement of two QDs could be achieved by adjusting the direct coupling strength of the QDs, the interaction of QDs with near-zero waveguide modes, interparticle distance of the QDs, total dissipation and detuning even when two QDs are resonant with the incident single plasmon. The discussed system with the flip-flop interaction provides us rich way to realize the quantum information processing such as quantum communication and quantum computation.

# 1. Introduction

During recent years, the control of the light at the nanoscale based on the nature of plasmons plays important roles in quantum information science, such as schemes for quantum cryptography, quantum teleportation, and quantum computation [1-4], however, its practical realization is still challenging because the requisite single-photon nonlinearities are generally very weak. Many theoretical [5-8] and experimental works [9-11] have extensively sought to achieve the strong coupling between the quantum emitter and single photon. Its fundamental concept is based on the coupling the qubits to a structured EM environment by using quantum emitter placed close to a plasmonic waveguide. The plasmonic waveguide with its radius of the optical subwavelength exhibits good confinement and guiding for the single photon (i.e., single plasmon). In this limit, the strong confinement results in reduction the effective mode volume $V_{eff}$ for the photons, thereby achieving a substantial increase in the coupling strength and an effective Purcell factor $P \equiv \Gamma_{pl} / \Gamma'$ can exceed $10^3$ in realistic systems according to the theoretical results [12], where $\Gamma_{pl}$ is the spontaneous emission rate into the surface plasmons(photons) and $\Gamma'$ describes contributions from both emission into free space and non-radiative emission via Ohmic losses in the conductor.

A single plasmon transport [13-16] and an adequate controlling way of quantum entanglement[17-19] based on the plasmonic waveguide provide us many promising applications for design of simple quantum logic gates. Two main ways have been proposed to investigate the single plasmon transport and entanglement of two qubits: optical cavities [20] with single mode or a collection of discrete modes and waveguides [21, 22] with continuum of modes. Especially, the real-space method [5] is investigated theoretically for single plasmon transport[13-16] and entanglement of QDs[18, 19], which indicates that the high spontaneous emission can be achieved for systems with a quantum dot doped inside a photonic crystal waveguide. Up to the present, they mainly focus on the cases where the quantum emitters are all the same and they also neglect the interaction between quantum emitters. However, the interaction of the quantum emitters plays important role for the transport of the single plasmon as well as the generation of the concurrence (i.e., entanglement) because it is relevant to the Förster resonance energy transfer (FRET) from donor QD to acceptor QD.

In the present work, we investigate theoretically the entanglement of two QDs by means of the concurrence via a real-space method, considering the flip-flop interaction. First, we obtain the dynamics equations of the possible-manufacturing system composed of two QDs and a plasmonic waveguide. After then, on the basis of the analytical solutions of the system, we find the possibilities of control of the entanglement by adjusting the direct coupling strength of the QDs, the interaction of QDs with near-zero waveguide modes, interparticle distance of the QDs, total dissipation and detuning even when two QDs are resonant with the incident single plasmon.

## 2. Model and dynamics equations

The system we consider here consists of two QDs with the flip-flop interaction and plasmonic waveguide as shown in Fig. 1, where the interparticle distance of the two QDs with excited state $|e\rangle$ and ground state $|g\rangle$ is $d$ and they are placed in the vicinity of a cylindrical plasmonic waveguide with its diameter of $D$. The source of the left incoming single plasmon (e.g., an additional QD) must be located on near the first QD to minimize the initial losses because the propagating single plasmon along the surface of the plasmonic waveguide inevitably occurs dissipation (such as Ohmic loss). The system is embedded in dielectric medium such as $SiO_2$ and the cut-off wavelength of a propagating single plasmon with near-zero mode index could be controlled by changing the diameter of the plasmonic waveguide, where its group velocity is varied according to the mode index in the plasmonic waveguide with a certain diameter. For example, the single plasmon with near-zero mode index can propagate in range of wavelength from visible 685 nm to near-infrared 920 nm when the waveguide diameter $D$ changes from 100 nm to 150 nm. And the velocity $v_g$ corresponding to the near-zero mode index $n$=0.022, 0.962 are $0.82\times10^7$m/s, $9.55\times10^7$m/s, respectively, in near the cut-off wavelength of the plasmonic waveguide with its diameter $D$=110nm.

We investigate the entanglement of the two QDs interacting with the propagating single plasmon by using real space approach, where entanglement of the QDs can be quantified by the concurrence $C$ [23, 24].

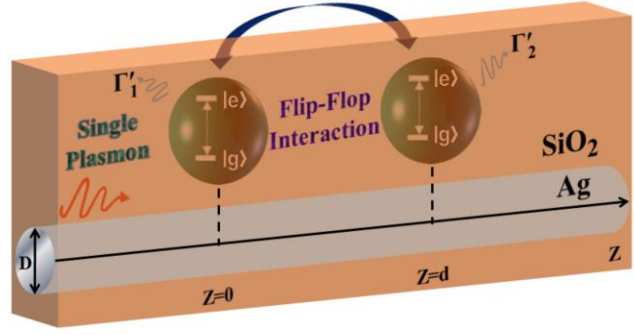

**Fig. 1** (Color online). The schematic diagram of a plasmonic waveguide coupled to two QDs with flip-flop interaction. $D$ is the diameter of the plasmonic waveguide and the interparticle distance of two-level QDs is $d$, where they are placed in the vicinity of the plasmonic waveguide and $\Gamma'_1(\Gamma'_2)$ is the total dissipation of the QD$_1$ (QD$_2$).

Under the rotating wave approximation, the Hamiltonian of the system in real space is given by [5]

$$H/\hbar = iv_g \int_{-\infty}^{\infty} dz [a_l^+(z)\partial_z a_l(z) - a_r^+(z)\partial_z a_r(z)] + \sum_{j=1}^{2}(\Omega_j - i\Gamma'_j/2)\sigma_j^+\sigma_j$$
$$+ \sum_{j=1}^{2}\sum_{\substack{i=1\\i\neq j}}^{2} f_{ji}\sigma_j^+\sigma_i + \sum_{j=1}^{2} g_j\{[a_r^+(z_j) + a_l^+(z_j)]\sigma_j + [a_r(z_j) + a_l(z_j)]\sigma_j^+\}, \quad (1)$$

where $a_r^+(z_j)$ ($a_l^+(z_j)$) is the bosonic operator creating a right-going (left-going) plasmon at position $z_j$ of the $j$th QD, $v_g$ is the group velocity of the surface plasmon with wavevector, $k$, so that $\omega_k$ is the frequency of the surface plasmon (i.e., $\omega_k = v_g|k|$). $\Omega_j$ are the transition energies of the $j$th QD, respectively, $\sigma_j^+ = |e\rangle_{jj}\langle g|$ ( $\sigma_j=|g\rangle_{jj}\langle e|$ ) is the raising (lowing) operators of the $j$th QD. The non-Hermitian term $\Gamma'_j$ is the total dissipation including the decay rate into free space and other dissipative channels, such as the Ohmic loss. We can include the Ohmic loss of plasmon during the scattering process in this way because a loss of the plasmon is equivalent to a loss of the excitation in the quantum dots. $g_j = (2\pi\hbar/\omega_k)^{1/2}\Omega_j \mathbf{D}_j \cdot \mathbf{e}_k$ is the coupling strength of the $j$th QD with the waveguide mode, $\mathbf{D}_j$ is the dipole moment of the $j$th QD, $\mathbf{e}_k$ is the polarization unit vector of the surface plasmon[5]. $f_{ji} = (\Gamma'_j/2)\cdot\beta e^{-l/(2L)}\sin(kd)$ describes the flip-flop interaction of two QDs, where L is the propagation length of the single plasmon[17].

Supposing a single plasmon incident from the left with energy $E_k = \hbar\omega_k$, then the eigenstate of the system, defined by $H|\psi_k\rangle = E_k|\psi_k\rangle$, can be written as follows:

$$|\psi_k\rangle = \int dz[\phi_{k,r}^+(z)a_r^+(z) + \phi_{k,l}^+(z)a_l^+(z)]|0,g_1,g_1\rangle + e_k^{(1)}|0,e_1,g_2\rangle + e_k^{(2)}|0,g_1,e_2\rangle, \quad (2)$$

where $|0, g_1, g_2\rangle$ denotes the vacuum state with zero plasmon and two QDs being unexcited, $|0, e_j, g_i\rangle$ denotes the vacuum field and only the $j$th QD in the excited state and $e_k^{(j)}$ is the probability amplitude of the $j$th QD in the excited state. $\Phi^+_{k,r}(z)$ ($\Phi^+_{k,l}(z)$) is the mode function of a right-going (a left-going) plasmon at position $z$.

For a plasmon incident from the left, the mode functions $\Phi^+_{k,r}(z)$ and $\Phi^+_{k,l}(z)$ take the forms: $\phi^+_{k,r}(z<0)=e^{ikz}$, $\phi^+_{k,r}(0<z<d)=t_1 e^{ik(z-d)}$, $\phi^+_{k,r}(z>d)=t_2 e^{ik(z-2d)}$, $\phi^+_{k,l}(z<0)=r_1 e^{-ikz}$, $\phi^+_{k,l}(0<z<d)=r_2 e^{-ik(z-d)}$, and $\phi^+_{k,l}(z>d)=0$, respectively. Here $t_j$ and $r_j$ ($j=1, 2$) are the transmission and reflection amplitudes at the place $z_j$, respectively. By substituting Eq. (2) into $H|\psi_k\rangle = E_k|\psi_k\rangle$ and taking the boundary conditions of the mode functions $t_{j-1}+r_j = t_j e^{-ikd} + r_{j+1}e^{ikd}$, where $t_0 = 1$, $r_3 = 0$ and $j = 1, 2$, into account, we obtain the dynamics equations of the system as shown Fig. 1, respectively, as $g_1(t_1 e^{-ikd}-1)+iJ_1 e_k^{(1)}=0$, $g_1(r_2 e^{ikd}-r_1)-iJ_1 e_k^{(1)}=0$, $g_1(1+r_1)+(\Delta_1-\Gamma'_1/2)e_k^{(1)}+f_{12}e_k^{(2)}=0$, $g_2(t_2 e^{-ikd}-t_1)+iJ_2 e_k^{(2)}=0$, $-g_2 r_2-iJ_2 e_k^{(2)}=0$, $g_2(t_1+r_2)+ (\Delta_2-\Gamma'_2/2)e_k^{(2)}+f_{21}e_k^{(1)}=0$, where $J_j = g^2_j/v_g$, $\Delta_j = \Omega_j - \omega_k$ ($j = 1, 2$).

## 3. Theoretical analysis and numerical results

In principle, the transmission and reflection spectra of a single plasmon can be observed by detectors placed at two ends of the plasmonic waveguide [23]. However, if there is no detection of the surface plasmon at both ends of the plasmonic waveguide, the state of the system is projected onto the excited states of two QDs, which means that create entanglement of two QDs. We can obtain the analytical solution of the above equation set in the steady state, from which we can investigate the entanglement degree of two QDs, that is, the concurrence $C$. For the system consisting of two QDs, concurrence $C$ takes the form of $C = \dfrac{2|e_k^{(1)}||e_k^{(2)}|}{|e_k^{(1)}|^2 + |e_k^{(2)}|^2}$. In all our calculations, we suppose that $f_{12} = f_{21} = f$ ($j = 1, 2$), and $D=110$ nm.

First of all, we investigate the dependence of the concurrence on the coupling strength between QDs and plasmonic waveguide, $g$, when two QDs with the same transition energy are resonant with the propagating single plasmon and $g_1 = g_2 = g$. In that case, assuming that there is no decay rate, we can obtain probability amplitude of the $j$th QD in the excited state, which is as follows.

$$\begin{cases} e_k^{(1)} = -\dfrac{ig\upsilon_g\left[\left(-1+e^{i2kd}\right)g^2 + ie^{ikd}f\upsilon_g\right]}{\left(-1+e^{i2kd}\right)g^4 + 2ie^{ikd}fg^2\upsilon_g - f^2\upsilon_g^2}, \\ e_k^{(2)} = -\dfrac{fg\upsilon_g^2}{\left(-1+e^{i2kd}\right)g^4 + 2ie^{ikd}fg^2\upsilon_g - f^2\upsilon_g^2}. \end{cases} \quad (3)$$

As you can see easily from Eq. (3), if there is no interparticle coupling effect between two QDs, entanglement of two QDs does not occur. But under the flip-flop interaction of two QDs we can control the concurrence by adjusting the parameters such as the coupling strength of the direct-coupled QDs and interaction between QDs and the plasmonic waveguide, group velocity of the propagating single plasmon and the interparticle distance of the QDs.

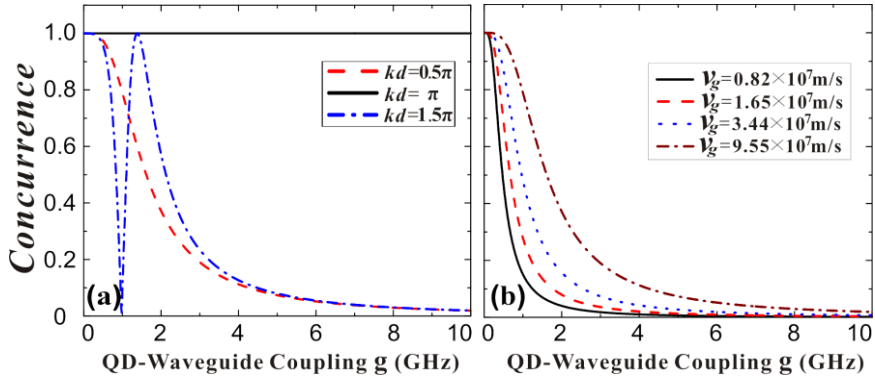

**Fig. 2** (Color online). Dependence of concurrence $C$ on the QD-waveguide coupling $g$ for $f$=20GHz. (a) $kd$=0.5π (solid line), π (dashed line), 1.5π (dash-dotted line) in case of $v_g$=9.55×10$^7$m/s (b) $v_g$=0.82×10$^7$m/s (solid line), $v_g$=1.65×10$^7$m/s (dashed line), $v_g$=3.44×10$^7$m/s (dotted line), $v_g$=9.55×10$^7$m/s (dash-dotted line) in case of $kd$=0.5π. Here two QDs have the same transition energies ($\Omega_1=\Omega_2$) and no detuning ($\Delta_j=0$).

From Fig. 2, we can find a graphic illustration of entanglement of two QDs in the presence of the flip-flop interaction as a function of coupling strength between QDs and waveguide when the frequencies of two QDs with the same transition energy are resonant with the frequency of single plasmon, where we set the interparticle coupling strength, $f$=20GHz. Fig. 2(a) shows the concurrence $C$ at near-zero mode index n=0.962 corresponding group velocity $v_g$=9.55×10$^7$m/s when $kd$=0.5π (solid line), π (dashed line), 1.5π (dash-dotted line), respectively. In case of $kd$=π, the concurrence $C$ has the maximum value as $C$=1 for any QD-waveguide coupling strength, which means that the amplitude $e_k^{(1)}$ is equal to $e_k^{(2)}$ or $-e_k^{(2)}$. In this case, the two QDs become triplet- or singlet-entangled. However, when the coupling QD-waveguide gets strong, the concurrence $C$ for $kd$=0.5π approaches from 1 to 0 and especially, the concurrence $C$ for $kd$=1.5π has a dip with its minimum value of 0. Fig. 2(b) shows the dependence of concurrence $C$ on the QD-waveguide coupling for $kd=0.5\pi$ when the waveguide mode index n=0.022, 0.164, 0.462, and 0.962. The corresponding group velocity $v_g$ are $0.82\times10^7$m/s, $1.65\times10^7$m/s, $3.44\times10^7$m/s and $9.55\times10^7$m/s, respectively, where concurrence $C$ goes from 1 to 0 more quickly when the waveguide mode index $n$ approaches to zero.

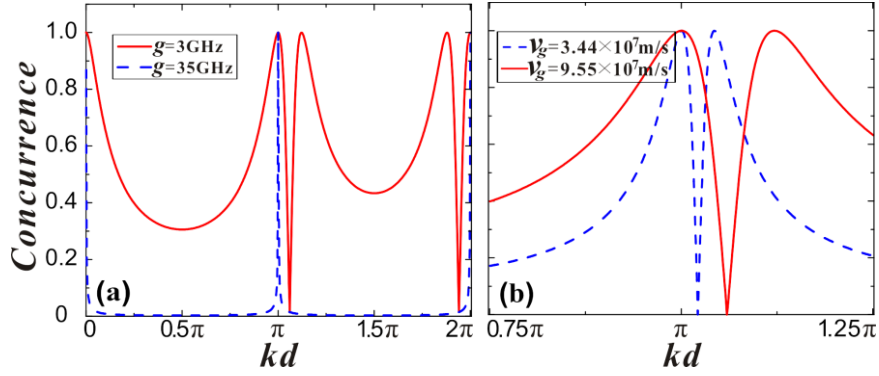

**Fig. 3** (Color online). Concurrence $C$ versus the phase difference between two QDs, $kd$ for $f$=35GHz (a) $g$=3GHz (solid line), 35GHz (dashed line) in case of $v_g$=9.55×10$^7$m/s (b) $v_g$=3.44×10$^7$m/s (dashed line), $v_g$=9.55×10$^7$m/s (solid line) in case of $g$=3GHz. Here two QDs have the same transition energies ($\Omega_1=\Omega_2$) and no detuning ($\Delta_j$ =0).

In Fig. 3 we discuss the dependence of concurrence $C$ on the phase difference between two QDs, $kd$, when the frequencies of two QDs with the same transition energy are resonant with the frequency of single plasmon, where we take the fixed interparticle coupling strength, $f$=35GHz. In principle, the phase difference between two QDs means the interparticle distance, which plays important role in not only the transport of a single plasmon but also entanglement of two QDs. Fig. 3(a) shows the influence of the phase difference $kd$ on the concurrence $C$ when the group velocity $v_g$ are $9.55\times10^7$m/s, corresponding to waveguide mode index $n$=0.962. As you can see from Fig. 3(a), the concurrence $C$ has the periodic property with its period of π. In particular, for the QD-waveguide coupling strength, $g$=3GHz, the concurrence $C$ has a tapering peak and maximum value 1 in phase $kd$=nπ (n=0, 1, 2, …), which is great useful to achieve high entanglement of two QDs and switch on or off. For the QD-waveguide coupling strength, $g$=35GHz the concurrence $C$ has double peaks around $kd$=nπ and its maximum values are 1. In Fig. 3(b) we show concurrence $C$ as a function of phase difference between QDs for $g$=35GHz, where it has double peaks and the space between two peaks for $v_g$=3.44×10$^7$m/s is smaller than for $v_g$=9.55×10$^7$m/s.

Next, regarding total dissipation, we consider the concurrence $C$ versus QD-waveguide coupling strength, $g$, when two QDs with the same transition energy are resonant with the propagating single plasmon and $g_1=g_2=g$. We can obtain probability amplitude of the $j$th QD in the excited state, which is as follows.

$$e_k^{(1)} = -\frac{2ig\upsilon_g\left[2\left(-1+e^{i2kL}\right)g^2 + \left(2ie^{ikL}f - \Gamma'\right)\upsilon_g\right]}{4\left(-1+e^{i2kL}\right)g^4 + 4ig^2\left(2e^{ikL}f + i\Gamma'\right)\upsilon_g - \left(4f^2 + \Gamma'^2\right)\upsilon_g^2},$$

$$e_k^{(2)} = \frac{2g\left(2f + ie^{ikL}\Gamma'\right)\upsilon_g^2}{4\left(-1+e^{i2kL}\right)g^4 + 4ig^2\left(2e^{ikL}f + i\Gamma'\right)\upsilon_g - \left(4f^2 + \Gamma'^2\right)\upsilon_g^2}.$$

(4)

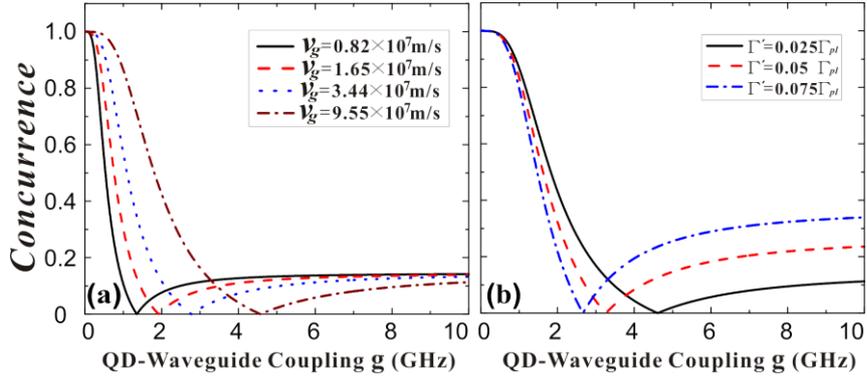

**Fig. 4** (Color online). Influence of the QD-waveguide coupling strength, $g$ on concurrence $C$. (a) $v_g=0.82\times10^7$m/s(solid line), $v_g=1.65\times10^7$m/s(dashed line), $v_g=3.44\times10^7$m/s(dotted line), $v_g=9.55\times10^7$m/s(dash-dotted line) for $\Gamma'=0.025\Gamma_{pl}$ (b) $\Gamma'=0.025\Gamma_{pl}$ (solid line), $\Gamma'=0.05\Gamma_{pl}$ (dashed line), $\Gamma'=0.075\Gamma_{pl}$ (dash-dotted line) for $v_g=9.55\times10^7$m/s, where $kd=0.5\pi$, $f=35$GHz, $\Gamma_{pl}=4\pi g^2/v_g$. Here two QDs have the same transition energies ($\Omega_1=\Omega_2$) and no detuning ($\Delta_j=0$).

Fig. 4 numerically displays the concurrence $C$ versus the QD-waveguide coupling strength $g$ when there exists the total dissipation of which unit is the decay rate into waveguide modes as $\Gamma_{pl}=4\pi g^2/v_g$, where two QDs have the same transition energies ($\Omega_1=\Omega_2$) and no detuning ($\Delta_j=0$) and we take $f=35$GHz, $kd=0.5\pi$. As shown in Fig. 4(a), When there exist total dissipation($\Gamma'=0.025\Gamma_{pl}$), the concurrence $C$ is gradually decreased to has minimum value 0, and then approaches to a certain value (not zero) of the concurrence for various group velocities, $v_g=0.82\times10^7$m/s, $1.65\times10^7$m/s, $3.44\times10^7$m/s, and $9.55\times10^7$m/s, respectively. In Fig. 4(a) the gradient of concurrence $C$ becomes bigger when the group velocity gets bigger. Fig. 4(b) shows the concurrence $C$ when total dissipation is $\Gamma'=0.025\Gamma_{pl}$, $0.05\Gamma_{pl}$, $0.075\Gamma_{pl}$, respectively, where the bigger total dissipation is, the more quickly the concurrence decrease to 0 and the bigger value it goes to when the QD-waveguide coupling gets strong. As you can see from Fig. 4, the existence of total dissipation get changed dramatically the concurrence in comparison with the case of no-decay rate, which has great worth to investigate the entanglement because the dissipation could not be avoidable to construct the realistic quantum system.

Now, we consider the detuning between QDs and the propagating single plasmon to obtain probability amplitude of the $j$th QD in the excited state, which is as follows;

$$e_k^{(1)} = \frac{g\upsilon_g\left[\left(-1+e^{i2kL}\right)g^2 + i\left(e^{ikL}f - \Delta\right)\upsilon_g\right]}{-i\left(-1+e^{i2kL}\right)g^4 + 2g^2\left(e^{ikL}f - \Delta\right)\upsilon_g + i\left(f^2 - \Delta^2\right)\upsilon_g^2},$$

$$e_k^{(2)} = \frac{g\left(f - e^{ikL}\Delta\right)\upsilon_g^2}{-i\left(-1+e^{i2kL}\right)g^4 + 2g^2\left(e^{ikL}f - \Delta\right)\upsilon_g + i\left(f^2 - \Delta^2\right)\upsilon_g^2}, \quad (5)$$

where we assume that two QDs has the same detuning with single plasmon and there does not exist total dissipation.

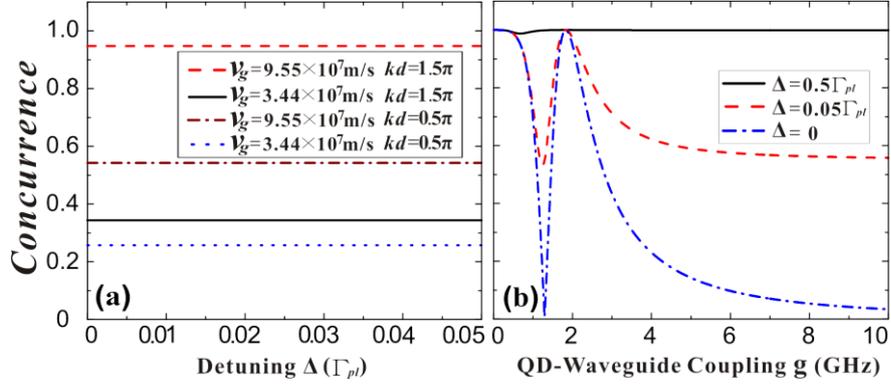

**Fig. 5** (Color online). Concurrence $C$ according to detuning in case of the same detuning of two QDs (i.e. $\Delta_j = \Delta \neq 0$). (a) $v_g=9.55\times10^7$m/s, $kd=1.5\pi$(dashed line), $v_g=3.44\times10^7$m/s, $kd=1.5\pi$(solid line), $v_g=3.44\times10^7$m/s, $kd=0.5\pi$(dash-dotted line), $v_g=9.55\times10^7$m/s, $kd=0.5\pi$(dotted line) for $g=2$GHz (b) $\Delta=0$(dash-dotted line), $\Delta=0.05\,\Gamma_{pl}$ (dashed line), $\Delta=0.5\,\Gamma_{pl}$ (solid line) for $kd=1.5\pi$, $v_g=9.55\times10^7$m/s, where $f=35$GHz, $\Gamma_{pl} = 4\pi g^2/v_g$.

Concurrence $C$ where two QDs has the same detuning and not zero is shown in Fig. 5. From Fig. 5(a) we can see the concurrence $C$ has the fixed value, not depending on detuning when each QD-waveguide coupling strength has the same value of $g=2$GHz and $f=35$GHz. Concurrence $C$ become higher when group velocity and interparticle distance get increased. It is important to maintain a fixed entanglement of two qubit in quantum information processing where QDs are generally used as qubit. These results on the concurrence of two QDs can be used in quantum teleportation and quantum computation. Fig. 5(b) show the concurrence $C$ for $kd=1.5\pi$ as function of the QD-waveguide coupling when each QD is resonant with the single plasmon, or not, where the dip between two peaks become shallower when the detuning gets increased. Moreover, when detuning $\Delta$ is $0.5\Gamma_{pl}$, the dip between two peaks disappears and the concurrence with maximum value of 1 is maintained for any QD-waveguide coupling strength.

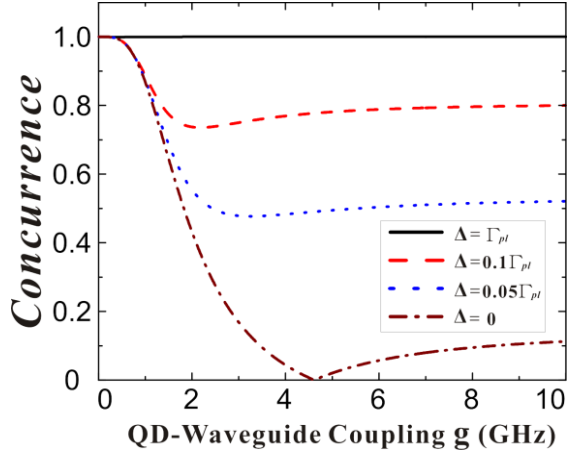

**Fig. 6** (Color online). Concurrence $C$ versus the QD-waveguide coupling strength, where we set $kd$=0.5$\pi$, $v_g$=9.55×10$^7$m/s, $f$=35GHz, $\Gamma = 0.025\Gamma_{pl}$ and $\Gamma_{pl} = 4\pi g^2/v_g$. $\Delta$=0(dash-dotted line), $\Delta$=0.05$\Gamma_{pl}$ (dotted line), $\Delta$=0.1$\Gamma_{pl}$ (dashed line), $\Delta$=$\Gamma_{pl}$ (solid line). In all case, two QDs have the same transition energy and the QD-waveguide coupling strength.

Next, we discuss the concurrence $C$ versus the QD-waveguide coupling strength when two QDs have the same transition energies and the QD-waveguide coupling strength, where $kd$=0.5$\pi$, $v_g$=9.55×10$^7$m/s, $f$=35GHz, $\Gamma = 0.025\Gamma_{pl}$, $\Gamma_{pl} = 4\pi g^2/v_g$ and $\Delta$=0(solid line), $\Delta$=0.05 $\Gamma_{pl}$ (dashed line), $\Delta$=0.1 $\Gamma_{pl}$ (dash-dotted line), $\Delta$=$\Gamma_{pl}$ (solid line). In Fig. 6, we can find concurrence $C$ has minimum value 0 in a certain QD-waveguide coupling strength when two QDs are resonant with the frequency of the single plasmon. However, as the detuning becomes large, the two QDs are kept high concurrence in wide range of the QD-waveguide coupling strength. In particular, concurrence $C$ does not get changed for any QD-waveguide coupling strength when $\Delta=\Gamma_{pl}$.

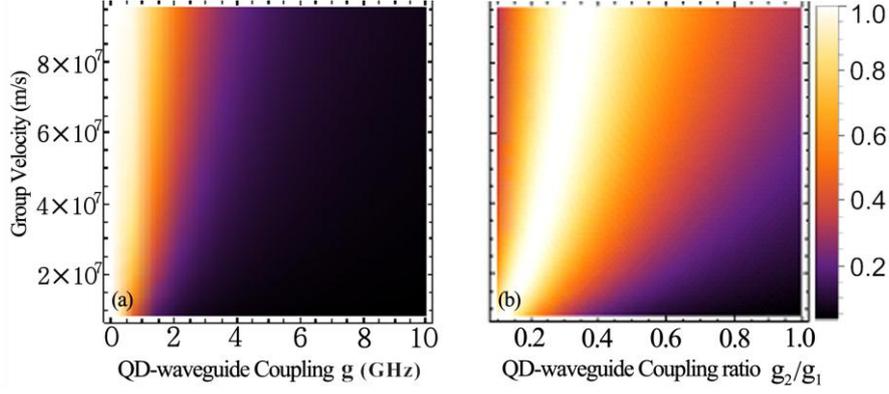

**Fig. 7** (Color online). Concurrence *C* versus the QD-waveguide coupling strength and the group velocity, where $kd=0.5\pi$, $f=35$GHz. (a) $g_1=g_2=g$, (b) $g_1 \neq g_2$, $g_1=3$GHz. Here two QDs have the same transition energies ($\Omega_1=\Omega_2$) and no detuning ($\Delta_j=0$).

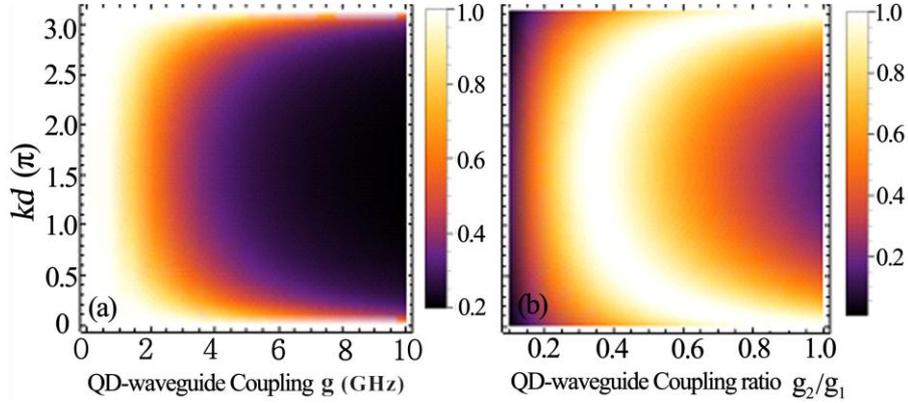

**Fig. 8** (Color online). Concurrence *C* versus the QD-waveguide coupling strength and interparticle distance of QDs, where $v_g=9.55\times10^7$m/s, $f=35$GHz, (a) $g_1=g_2=g$, (b) $g_1 \neq g_2$, $g_1=3$GHz.. Here the two QDs have the same transition energies ($\Omega_1=\Omega_2$) and no detuning ($\Delta_j=0$).

In the above discussions, we investigated the entanglement of two QDs, assuming that each QD coupled to plasmonic waveguide has the same coupling strength. In Fig. 7 and Fig. 8 we can find the influence of difference of QD-waveguide coupling on the concurrence *C*, where we set $kd=0.5\pi$, $f=35$GHz. Fig. 7(a) shows the concurrence *C* versus the QD-waveguide coupling strength and the group velocity when each QD coupled to plasmonic waveguide has the same coupling strength and Fig. 7(b) shows the concurrence *C* as functions of the group velocity and the ratio of the QD-waveguide coupling strengths when $g_1=3$GHz. As you can see easily from Fig. 7, high entanglement could be achieved when each QD-waveguide coupling strength is not equal to each other, which is easy to realize in practice by adjusting the distance between the QD and waveguide. Fig. 8(a) shows the concurrence *C* versus the QD-waveguide coupling strength and interparticle distance of QDs when each QD coupled to plasmonic waveguide has the same coupling strength and Fig. 8(b) shows the concurrence *C* as functions of the interparticle distance of QDs and the ratio of the QD-waveguide coupling strengths when $g_1$=3GHz. As shown in Fig. 8, high entanglement could be achieved when each QD-waveguide coupling strength does not equal each other, which is easy to realize in practice by adjusting the interparticle distance of QDs.

**4. Conclusion**

In conclusion, we discussed theoretically the entanglement of two QDs coupled to a metallic nanowaveguide under the flip-flop interaction, considering the dependence of concurrence on various physical parameters. We showed that the entanglement of two QDs could be achieved by adjusting the direct coupling strength of the QDs, the interaction of QDs with near-zero waveguide mode, interparticle distance of the QDs, total dissipation and detuning. Setting the above parameters properly results in the switching between high and low entanglement even when the two QDs are resonant with the incident single plasmon. In particular, our results show that the flip-flop interaction plays important role to achieve high entanglement of two QDs, which provides us rich way to realize the quantum information processing such as quantum communication and quantum computation.

**Acknowledgments.** This work was supported by the National Program on Key Science Research of DPR of Korea (Grant No. 131-00). This work was also supported by the National Natural Science Foundation of China (NSFC) under Grants Nos. 11404410, 11504105 and 11504434.